\documentstyle[12pt]{article}

\title
{Quantization of systems with time-dependent constraints. Example of
relativistic particle in  plane wave}

\author
{S.P.Gavrilov\\
Department of General Physics\\
Tomsk State Pedagogical Institute\\
Tomsk 634041\\
Russia
\and
D.M.Gitman \\
Instituto de F\'{\i}sica\\
Departamento de F\'{\i}sica Matem\'atica\\
Universidade de S\~ao Paulo, Caixa Postal 20516 \\
01498-S\~ao Paulo, S.P.\\
Brazil}

\begin{document}
\maketitle
\begin{abstract}
A modification of the canonical quantization procedure for systems with
time-dependent second-class constraints is discussed and applied to the
quantization of the relativistic particle in a plane wave. The time dependence
of constraints appears in the problem in two ways. The Lagrangian depends on
time explicitly by origin, and a special time-dependent gauge is used. Two
possible approaches to the quantization are demonstrated in this case. One is
to solve directly a system of operator equations, proposed by Tyutin and one of
the authors (Gitman) as a generalization of  Dirac canonical
quantization in nonstationary
case, and another to find first a canonical transformation, which makes it
possible to discribe the dynamics in the physical sactor by means of some
effective Hamiltonian. Quantum mechanics
constructed in both cases proves to be equivalent to Klein-Gordon theory of the
relativistic particle in a plane wave. The general conditions of unitarity
of the dynamics in the physical sector are discussed.
\end{abstract}

\section{Introduction}
 In the relativistic partical theories, string theories, theories of gravity
and so on, these appear sometimes constraints which depend on time manifestly.
The causes are both an explicit time dependence of a Lagrangian, theories in
external nonstationary fields may be an example of that, and necessity
sometimes to impose gauge conditions which depend on time. The rules of
canonical quantization of systems with second-class constraints \cite{ab1},
which do not
depend on time, have to be changed in that case. The corresponding
modification was proposed in \cite{ab2}. The canonical operator quantization
implies in this case the solution of both commutation relations with a Dirac
bracket, equations of constraints and, besides, some differential equations,
which determine the time evolution of operators due to the time-dependence
of constraints. This procedure is brifly discribed in Sect.2. In general
case the whole time-evolution is not unitary one and can not  be reduced to
 the solution of the Schr\"odinger equation only with some Hamiltonian.
However, there exist cases
when one can  make constraints time-independent by means of a classical
canonical transformation and then use the ordinary rules of quantization.
Particularly, in
such a way an operator quantization of the relativistic
particle in the constant magnetic field and a chronological gauge was
fulfiled  \cite{ab3}.

In this paper (Sect.3) we consider the  quantization
 of the
relativistic spinless particle in the external field of a plane wave, which is
instructive as a demonstration how does the general theory works and at the
same time is new and interesting problem itself.
 The time
dependence of constraints appears here in two possible ways, the Lagrangian
depends on time by origin and we use a special time-dependent  gauge
condition. We demonstrate here both of above mentioned possibilities in
the quantization of systems with time-dependent constraints. It turns out
that in spite of the dynamics is not unitary one in the whole, it is unitary
in the physical sector. In Sect.4 we consider such a situation in general
terms and discuss the problem of the canonical quantization of a relativistic
particle in different backgrounds.

\section{Canonical quantization of theories with time-dependent second-class
constraints}

Here, we briefly describe the modification of the Dirac brackets method of
quantization for time-dependent second class constraints \cite{ab2}.

Let we have a theory in Hamiltonian formulation with second-class constraints
$\Phi(\eta,t)=0,\; \eta=(q,p)$,   which can explicitly depend on time $t$. Then
the equation of motion of such a system may be written in the usual form, if
one
 formally introduces a momentum  $\epsilon$   conjugated to the time $t$, and
defines the Poisson bracket in the extended space of canonical variables
$(q,p;t,\epsilon)=(\eta;t,\epsilon)$,

\begin{equation}
\dot{\eta}=\{\eta,H + \epsilon\}_{D(\Phi)},\;  \Phi(\eta, t) =0,\label{e1}
\end{equation}

\noindent where H is a Hamiltonian of the system, and   $\{A,B\}_{D(\Phi)}$
is the denotation
for the Dirac bracket with respect to a system of  second-class
constraints $\Phi$. The Poisson bracket, wherever encountered, is henceforth
understood as one in such above mentioned extended space. The total derivative
of an arbitrary function   $A(\eta,t)$,   with allowance made for the equations
(\ref{e1}), has the form

\[
\frac{dA}{dt}=\{A,H+\epsilon\}_{D(\Phi)} \;.
\]

The quantization procedure in "Schr\"odinger" picture can be formulated in
that case as follows. The variables  $\eta$   of the theory are assigned the
operators  $ \hat{\eta} $,  which satisfy the equal-time commutation relations
($[,\}$ is a denotation for the generalized commutator,
commutator or anticommutator depending on parities of variables),

\begin{equation}\label{e2}
[\hat{\eta},\hat{\eta}'\}=i\{\eta,\eta'\}_{D(\Phi)}|_{\eta=
\hat{\eta}},
\end{equation}

\noindent the constraints equations

\begin{equation}  \label{e3}
\Phi(\hat{\eta},t)=0   ,
\end{equation}

\noindent and equations of evolution (We disregard problems connected
with operators ordering),

\begin{eqnarray}
\dot{\hat{\eta}}
&=&\{\eta,\epsilon\}_{D(\Phi)}|_{\eta=\hat{\eta}} \nonumber   \\
&=&-\{\eta,\Phi_{l}\}\{\Phi,\Phi\}^{-1}_{ll'}\frac{\partial
\Phi_{l'}}{\partial t}|_{\eta=\hat{\eta}}   \;  .\label{e4}
\end{eqnarray}

\noindent To each physical quantity A given in the Hamiltonian formalism by the
 function
$A(\eta,t)$, we assign a "Schr\"odinger" operator  $\hat{A}$ by the rule
$\hat{A} = A(\hat{\eta},t)$; in the same manner we construct the quantum
Hamiltonian  $\hat{H}$, according to the classical Hamiltonian $ H(\eta,t)$.
The time evolution of the state vector  $\Psi$ in the"Schr\"odinger" picture
is determined by the Schr\"odinger equation

\begin{equation} \label{e5}
i\frac{\partial\Psi }{\partial t}=\hat{H}\Psi , \;\hat{H}=H(\hat{\eta},t) \;.
\end{equation}

\noindent From (\ref{e4}) it follows, in particular, that

\begin{equation}   \label{e6}
\frac{d\hat{A}}{dt}=\{A,\epsilon\}_{D(\Phi)}|_{\eta=\hat{\eta}},
\end{equation}

\noindent and, as consequence of (\ref{e2}), for arbitrary "Schr\"odinger"
operators $\hat{A}, \hat{B}$, we have

\begin{equation}  \label{e7}
[\hat{A},\hat{B}\}=i\{A,B\}_{D(\Phi)}|_{\eta=\hat{\eta}}.
\end{equation}

It is possible to see that  quantum theories, which correspond to
different initial data for the equation (\ref{e4}), are equivalent.

One can to adduce some arguments in favor of the proposed quantization
procedure. For instance, to check that the correspondence principle between
classical and quantum equations of motion holds true in this procedure. To this
end we pass over to the Heisenberg representation, whose operators $\check{
\eta}$ are related to the operators  $\hat{\eta}$  as  $\check{\eta} = U^{-1}
\hat{\eta} U $,
where $ U $ is the operator of the evolution of the Schr\"odinger equation,
$ i\partial U / \partial t = \hat{H} U ,\;  U|_{t=0} = 1 $.  Heisenberg
operators  $\check{A}$  of an arbitrary physical quantity $A$  are constructed
  from
the corresponding  "Schr\"odinger" operator  $ \hat{A} $  in the same manner
$\check{A} = U^{-1} \hat{A} U $.  One can find the total time derivative of the
Heisenberg operator   $\check{A}$. Making use of (\ref{e6}), we have

\begin{equation}
\frac{d\check{A}}{dt}=U^{-1}\left[-i[\hat{A},\hat{H}\}+\{A,\epsilon\}
_{D(\Phi)}|_{\eta=\hat{\eta}}\right]U .\label{e8}
\end{equation}

\noindent Taking into account eq.(\ref{e7}) and the connection between the
operators $\hat{\eta}$
and $\check{\eta}$,  we derive the equation for the Heisenbert operator
$\check{A}$:

\begin{equation}
\frac{d\check{A}}{dt}=\{A,H+\epsilon\}_{D(\Phi)}|_{\eta=\check{
\eta}}\;,\label{e9}
\end{equation}

\noindent which coincides in form with the classical equation of motion .

\noindent It follows from (\ref{e9}) that the Heisenberg operators
$\check{\eta}$  also satisfy the equation

\begin{equation}    \label{e10}
\dot{\check{\eta}}=\{\eta,H+\epsilon\}_{D(\Phi)}|_{\eta=\check{\eta}}.
\end{equation}

\noindent Besides, one can easy verify, that  the equal-time
 relations hold  for these operators,

\begin{equation}   \label{e11}
[\check{\eta},\check{\eta}'\}=i\{\eta,\eta'\}_{D(\Phi)}|_{\eta=
\check{\eta}},\;\; \Phi(\check{\eta},t)=0
\end{equation}

\noindent The relations, together with (\ref{e10}), may be regarded as a
prescription of the quantization in the Heisenberg picture for theories with
time-dependent second-class constraints.

Note, that time dependence of Heisenberg operators in the theories, being
considered, is not unitary in  general case. In other words, no such
("Hamiltonian") operator exists, whose commutator with  a physical quantity
would give its total time derivative. This is explained by the existence of two
 factors which determin the time evolution of the Heisenberg operator. The
first
 one is a unitary evolution of the state vector in the "Schr\"odinger"
picture, while the second one is the time variation of "Schr\"odinger"
operators $\hat{\eta}$,  which in the general case is of nonunitary character.
The existence of these two factors is connected with the division of the right-
hand side of (\ref{e8}) into two summands. Physically, this is explained by
the fact
that the dynamics develops on a surface which itself changes with time; in the
 general
case, not in an unitary way.

\section{Quantization of the relativistic particle in a plane wave}

Let us consider a spinless particle in the external electromagnetic field of a
plane wave directed along the axis  $ x^{3} $.  The reparametrization invariant
action has  in that case the form

\begin{eqnarray}
S&=&-\int\left(m\sqrt{\dot{x}^{2}}+e\dot{x}A\right)d\tau \label{e12}\\
&=&-\int\left[m\sqrt{2\dot{x}_{+}\dot{x}_{-}-(\dot{x}_{\bot})^{2}}+e\dot{x}^
{a}A_{a}(x_{-})\right]d\tau, \nonumber \\
A_{\mu}&=&(0,A^{a},0),\; x_{\pm}=(x^{0}\pm x^{3}/\sqrt{2},\; x_{\bot}=(x^{a}),
\;a=1,2.\nonumber
\end{eqnarray}

Introducing the momenta

\begin{eqnarray}
\pi_{\pm}&=&\frac{\partial L}{\partial \dot{x}_{\pm}}=-\frac{m\dot{x}_{\mp}}
{\sqrt{2\dot{x}_{+}\dot{x}_{-}-(\dot{x}_{\bot})^{2}}},\label{e13} \\
\pi_{a}&=&\frac{\partial L}{\partial\dot{x}^{a}}=\frac{m\dot{x}^{a}}{\sqrt{
2\dot{x}_{+}\dot{x}_{-}-(\dot{x}_{\bot})^{2}}}-eA_{a}(x_{-}),\nonumber
\end{eqnarray}

\noindent one can see that a primary constraint exists,

\begin{equation}
\Phi^{(1)}=\pi_{-}-\frac{\left[\pi_{a}+eA_{a}(x_{-})\right]^{2}+m^{2}}
{2\pi_{+}}=0, \;\pi_{\pm}\neq 0. \label{e14}
\end{equation}

\noindent Going over to the Hamiltonian formulation, one has to express
velocities via momenta, by means of eq. (\ref{e13}). In the case of
consideration,
which is singular one, the velocities  $ \dot{x}_{+} $ and  $ \dot{x}_{\bot}$
  can only  be expressed in such a way:

\begin{equation}         \label{e15}
\dot{x}_{+}=\frac{\pi_{-}}{\pi_{+}}\dot{x}_{-},\;
\dot{x}_{a}=-\frac{(\pi_{a}+eA_{a})}{\pi_{+}}\dot{x}_{-}\;.
\end{equation}

\noindent Here, the velocity  $\dot{x}_{-}$  is primary unexpressible one.
Nevertheless, it follows from the eq.(\ref{e13}) that

\begin{equation}    \label{e16}
{\rm sign}\,\dot{x}_{-}=-{\rm sign}\,\pi_{+}=\zeta.
\end{equation}

\noindent Thus, in fact, the unexpressible is only the modulus of the
velocity   $\dot{x}_{-}$. In keeping with general recipes \cite{ab1,ab2}, we
construct a Hamiltonian   $ H^{(1)} $  by substituting in the expression
$\pi_{+}\dot{x}_{+} + \pi_{-}\dot{x}_{-} + \pi_{a}\dot{x}^{a} - L $,   the
velocities $\dot{x}_{+}, \dot{x}^{a}$   and ${\rm sign}\,\dot{x}_{-}$,
according to (\ref{e15}), (\ref{e16}). As a result, we obtain

\begin{equation}  \label{e17}
H^{(1)}=2\lambda\zeta\Phi^{(1)},\;\lambda=|\dot{x}_{-}|.
\end{equation}

\noindent One can readily make sure that the Hamiltonian equations with the
Hamiltonian (\ref{e17}) and primary constraint (\ref{e14}) are equivalent to
the Lagrange equations of motion. In particular,

\begin{equation}
\dot{\pi}_{+}=\dot{\pi}_{a}=0,\;\dot{\pi}_{-}=-e\dot{x}^{a}A_{a}',\;A_{a}'=
\frac{\partial A_{a}}{\partial x_{-}}.  \label{e18}
\end{equation}

The Hamiltonian (\ref{e17}) is equal to zero on the constraints surface,
because of the reparametrization invariance of the action. There are not more
constraints here, so we have only one first-class constraint (\ref{e14}). The
theory is degenerate, it is impossible to find   $ \lambda $   in the Dirac
procedure. We are following the canonical way and impose the gauge condition
similar to the one of the work \cite{ab3},

\begin{equation}
\Phi^{G}=x_{-}-\zeta\tau=0.   \label{e19}
\end{equation}

\noindent Considering the consequences of the conservation of this gauge in
the time  $\tau$ on the constraints surface,

\[
\frac{d\Phi^{G}}{d\tau}=\frac{\partial\Phi^{G}}{\partial\tau}+\{\Phi^{G},
H^{(1)}\}=\zeta(2\lambda-1)=0,
\]

\noindent we obtain   $\lambda = \frac{1}{2} $. No other constrains arise.The
complete set of constraints, in the gauge of consideration,   $\Phi = (\Phi^{G}
, \Phi^{(1)}) $, is of second-class one, and depends on time   $\tau$.

We consider here two ways of quantization. On the firt way (item a) we strictly
 follow the procedure described in the Sect. 2. On the second one (item b) we
use a canonical transformation, which, in combination with the first method,
allows one to reach the result more simple.

a) In the case of consideration, when the Hamiltonian H equals zero, the
``Schr\"odinger'' quantization  (\ref{e2}) - (\ref{e4})  coinside with the
Heisenberg one (\ref{e10}), (\ref{e11}).
The state vectors do not depend on time   $\tau$. All the time dependence is
due
 to the one of operators, according the equation (\ref{e2}) - (\ref{e4}) or
(\ref{e10}), (\ref{e11}). We
will denote operators which correspond to the variables $ \eta =
(x_{\pm},\pi_{\pm},x^{a}, \pi_{a}) $   as Heisenberg operators  $\check{\eta} $
. The denotation  $\hat{\eta} $  we will use in this item for operators in some
 special representation.

To write the equations of quantization (\ref{e2})-(\ref{e4}), we have to take
into account the following Dirac brackets:

\begin{eqnarray}
\{x_{+},\pi_{+}\}_{D(\Phi)}&=&1,\; \{x^{a},\pi_{b}\}_{D(\Phi)}=\delta_{b}^{a},
\label{e20} \\
\{x_{+},\pi_{-}\}_{D(\Phi)}&=&-\{x_{+},\Phi_{2}\},\;\{x^{a},\pi_{-}\}_{D(\Phi)}
=-\{x^{a},\Phi_{2}\}. \label{e21}
\end{eqnarray}

\noindent The Dirac brackets between the rest canonical variables are zero. The
 matrix, inverse to the one   $ \parallel \{\Phi_{l}, \Phi_{l'}\}\parallel  $,
 is

\[
\parallel \{ \Phi,\Phi\}^{-1} \parallel=\left(
\begin{array}{cc}
0&-1\\
1& 0
\end{array}
\right).
\]

\noindent So we can write the equal-time commutation relations (\ref{e2}) for
the
 operators of the independent canonical variables    $ x_{+}, \pi_{+}, x^{a},
\pi_{a},$

\begin{equation}
[\check{x}_{+},\check{\pi}_{+}]=i,\;[\check{x}^{a},\check{\pi}_{b}]=i
\delta^{a}_{b} \;,
\label{e22}
\end{equation}
\begin{eqnarray}
\left[\check{x}_{+},\check{\pi}_{-}\right]&=&i\{x_{+},\pi_{-}\}_{D(\Phi)}|
_{\eta=\check{\eta}}=-i\{x_{+},\Phi_{2}\}|_{\eta=\check{\eta}} \;,\nonumber \\
\left[\check{x}^{a},\check{\pi}_{-}\right]&=&i\{x^{a},\pi_{-}\}_{D(\Phi)}|
_{\eta=\check{\eta}}=-i\{x^{a},\Phi_{2}\}|_{\eta=\check{\eta}} \;.\label{e23}
\end{eqnarray}

\noindent The operator equations of constraints (\ref{e3}) are

\begin{equation}
\check{\Phi}_{1}=\check{x}_{-}-\check{\zeta}\tau=0,\;\check{\Phi}_{2}=
\check{\pi}_{-}-\frac{\left[\check{\pi}_{a}+eA_{a}(\check{x}_{-})\right]^{2}
+m^{2}}{2\check{\pi}_{+}}.  \label{e24}
\end{equation}

\noindent And the equations of evolution in time (\ref{e4}) look so

\begin{eqnarray}
\dot{\check{x}}_{+}&=&\{x_{+},\Phi_{2}\}\zeta|_{\eta=\check{\eta}},
\;\dot{\check{\pi}}_{+}=0, \label{e25} \\
\dot{\check{x}}^{a}&=&\{x^{a},\Phi_{2}\}\zeta|_{\eta=\check{\eta}},
\;\dot{\check{\pi}}=0, \nonumber \\
\dot{\check{x}}_{-}&=&\{x_{-},\Phi_{2}\}\zeta|_{\eta=\check{\eta}}
=\check{\zeta}, \nonumber  \\
\dot{\check{\pi}}_{-}&=&\{\pi_{-},\Phi_{2}\}\zeta|_{\eta=\check
{\eta}}=\frac{\partial}{\partial x_{-}}\left[\frac{\left(\pi_{a}+eA_{a}(x_{-})
\right)^{2}+m^{2}}{2\pi_{+}}\right]\zeta|_{\eta=\check{\eta}} \nonumber \\
&=&\frac{d}{d\tau}\left[\frac{\left(\check{\pi}_{a}+eA_{a}(\check{x}_{-})
\right)^{2}+m^{2}}{2\check{\pi}_{+}}\right]. \nonumber
\end{eqnarray}

\noindent The two last equations of (\ref{e25}) are, merely, the sequences of
the eq.
(\ref{e24}) of constraints. Taking into account eq. (\ref{e23}) and (\ref{e24})
, we can rewrite the equations of the evolution for the independent variables
in the form

\begin{eqnarray}
\dot{\check{x}}_{+}&=&i[\check{x}_{+},\check{\pi}_{-}\check{\zeta}]=-i[\check
{x}_{+},\check{H}], \label{e26} \\
\dot{\check{x}}^{a}&=&i[\check{x}^{a},\check{\pi}_{-}\check{\zeta}]=-i[\check
{x}^{a},\check{H}], \nonumber \\
\dot{\check{\pi}}_{+}&=&-i[\check{\pi}_{+},\check{H}]=0,\; \dot{\check{\pi}}_
{a}=-i[\check{\pi}_{a},\check{H}]=0, \nonumber
\end{eqnarray}
where
\[
\check{H}=\frac{\left[\check{\pi}_{a}+eA_{a}(\check{\zeta}\tau)\right]^{2}
+m^{2}}{|\check{\pi}_{+}|}.
\]

\noindent We suppose that  $\check{\pi}_{+}=-\check{\zeta}|\check{\pi}_{+}|$,
 like in classical theory. The difficulties with definition of such kind
 operators as $\check{\zeta}=-{\rm sign}\,\check{\pi}_{+},\;|\check{\pi}_{+}|$
and
$|\check{\pi}_{+}|^{-1}$, we are going to avoid,
 working in the representation of eigen functions of the operator $\check{\pi}
_{+}$.

As we mentioned in Sect.2, the time evolution of Heisenberg operators is not
unitary in general case. In the case of consideration, it is unitary, but
only for
the operators  $\check{x}_{+}, \check{x}^{a}, \check{\pi}_{+}, \check{\pi}_{a}
$, i.e. in the physical sector.

Let us go over from the picture in question, where state vectors do not
depend on time, and the operators depend on time according the equations
(\ref{e25}),(\ref{e26}), to another one (the corresponding operators will be
denoted as $ \hat{\eta} $)   connected with the former by quantum canonical
transformation    $ \check{\eta}=U^{-1}(\tau)\hat{\eta}U(\tau),\; U^{+}(\tau)=
 U^{-1}(\tau),\; U(0)=1 $,  so that, the operators of the independent
variables   $ \hat{x}_{+}, \hat{x}_{a},\hat{\pi}_{+},\\ \hat{\pi}_{a}$   in new
representation do not depend on time. To this end we have to choose the
operator  $ U(\tau)$  as the solution of the equation
$i\partial U/\partial\tau = \hat{H}U,\; U(0) = 1$   where

\begin{equation}    \label{e27}
\hat{H}=\frac{\left[\hat{\pi}_{a}+eA_{a}(\hat{\zeta}\tau)\right]^{2}+m^{2}}
{|\hat{\pi}_{+}|}.
\end{equation}

\noindent In the new representation, state vectors obey the equation of motion

\begin{equation}\label{e28}
i\frac{\partial \Psi}{\partial \tau} = \hat{H}\Psi,
\end{equation}

\noindent and the operators  $ \hat{x}_{+}, \hat{\pi}_{+}, \hat{x}^{a},
\hat{\pi}_{a} $   do not depend on time and have, according to (\ref{e22}),
(\ref{e26}), the canonical equal-time commutation relations

\begin{equation}\label{e29}
[\hat{x}_{+},\hat{\pi}_{+}]=i,\;[\hat{x}^{a},\hat{\pi}_{b}]=i\delta^{a}_{b}.
\end{equation}

We realize the commutation relations (\ref{e29}) in the space of function,
which depend on variables   $ \pi_{+}, x_{\bot}=(x^{a}) $,   as

\begin{equation}
\hat{\pi}_{+}=\pi_{+},\;\hat{x}_{+}=i\frac{\partial}{\partial\pi_{+}},\;
\hat{x}^{a}=x^{a},\;\hat{\pi}_{a}=-i\frac{\partial}{\partial x^{a}},\label
{e30}
\end{equation}

\noindent (in particular, $\hat{\zeta}=-{\rm sign}\,\pi_{+},\;|\hat{\pi}_{+}|=
|\pi_{+}|$) and by analogy with the classical theory, introduce the variable
 $ x_{-}$, which is defined as $x_{-}=\zeta\tau $ for eigen functions of the
operator  $\hat{\zeta}$  with the eigenvalue $\zeta$. Then the eq. (\ref{e28})
with the Hamiltonian (\ref{e27}) takes the form

\begin{equation}   \label{e31}
i\frac{\partial\Psi}{\partial x_{-}}=-\frac{\left[-i\frac{\partial}{\partial
 x^{a}}-eA^{a}(x_{-})\right]^{2}+m^{2}}{2\pi_{+}}\Psi.
\end{equation}

It is easy to verify that (\ref{e31}) is the Klein-Gordon equation. Indeed, the
Klein-Gordon equation

\begin{equation}   \label{e32}
({\cal P}^{2}-m^{2})\psi(x)=0,\;{\cal P}_{\mu}=i\partial_{\mu}-eA_{\mu}(x),
\end{equation}

\noindent in the external field of a plane wave (\ref{e12}), being written in
the light cone variables   $ (x_{\pm}, x_{\bot} = (x^{a}))$, has the form

\begin{equation} \label{e33}
\left(2\frac{\partial^{2}}{\partial x_{-}\partial x_{+}}+\left[-i\frac{
\partial}{\partial x^{a}}-eA^{a}(x_{-})\right]^{2}+m^{2}\right)\psi=0.
\end{equation}

\noindent Decomposing the  $\psi$ - function from (\ref{e33}) in the Fourier
integral with respect to the variable $x_{+}$,

\[
\psi(x_{-},x_{+},x_{\bot})=\frac{1}{\sqrt{2\pi}}\int^{\infty}_{-\infty}
e^{i\pi_{+}x_{+}}\Psi(x_{-},\pi_{+},x_{\bot})d\pi_{+} \;,
\]

\noindent we get from (\ref{e33}) the equation (\ref{e31}) for the function $
\Psi (x_{-},\pi_{+},x_{\bot})$.

The variable $\zeta$ is discrete and describes, as in case of the relativistic
 particle in the magnetic field \cite{ab3}, particle $\zeta=+$ and antiparticle
 $\zeta=-$. The demonstration of the statement on the classical level is quite
similar to that of the above mentioned work. To do that on the quantum level,
we have to interpret a function   $ \Psi(x_{-}, \pi_{+}, x_{\bot}) $  with
$\pi_{+} < 0,\;\zeta=+$,   as a wave function of a paticle, and a function
$\Psi^{*}(x_{-},\pi_{+},x_{\bot})$ with $\pi_{+}>0,\; \zeta=-$,
as the wave function of an antiparticle ( see also \cite{ab4} ),

\begin{eqnarray*}
\Psi_{+}&=&\Psi(x_{-},\pi_{+},x_{\bot}),\;\zeta=+ \;,\\
\Psi_{-}&=&\Psi^{*}(x_{-},\pi_{+},x_{\bot}),\;\zeta=- \;.
\end{eqnarray*}

\noindent Then, it follows from eq. (\ref{e31}),

\begin{eqnarray}
i\frac{\partial\Psi_{\zeta}}{\partial x_{-}}&=&\frac{\left[\hat{\vec{p}}_
{\bot}-g\vec{A}_{\bot}(x_{-})\right]^{2}+m^{2}}{2|\pi_{+}|}\Psi_{\zeta},
\label{e34} \\
\hat{\vec{p}}_{\bot}&=&-i\frac{\partial}{\partial\vec{x}_{\bot}},\;
\vec{A}_{\bot}(x_{-})=\left(A^{a}(x_{-})\right),\;g=\zeta e,\;\zeta=\pm.
\nonumber
\end{eqnarray}

\noindent Thus, the eq.(\ref{e34}) can be interpreted as two Schr\"odinger (in
 time $x_{-}$) equations with positive defined Hamiltonians, one for particle
with charge e, another for antiparticle with charge -e.

One can also prove, by analogy with the work \cite{ab3}, that the quantum
mechanics constructed is fully equivalent to the one-particle sector of the
quantum theory of the charged scalar field in the external electromagnetic
field of a plane wave. This statement can be also confirmed if one constructs
the
causal propagator $D^{c}(x,y)$ in Feynman spirit \cite{ab5} as the sum over
a complete set of the solutions of the equations (\ref{e34}), with $\zeta=
+$ at $x^{0}>y^{0}$, and with $\zeta=-$ at $x^{0}<y^{0}$. Such a calculation,
made in connection with some other problems, one can find in \cite{ab6,ab7} .
Solutions describing particles and antiparticles, were classified in that
calculation namely according the quantum number $\zeta$, which we have used
in the course of the quantization.

b) Here we point out an alternative way of quantization. Namely, already on the
 classical level, one can make the canonical transformation from the variables
 $\eta$ to the new ones  $\eta'$. The corresponding generating function has
the form

\begin{equation}
W=x_{+}\pi_{+}'+x_{-}\pi_{-}'+x^{a}\pi_{a}'+|\pi_{-}'|\tau,\label{e35}
\end{equation}

\noindent so that

\begin{eqnarray}
x_{-}'&=&x_{-}+{\rm sign}\,\pi_{-}\tau,\;x_{+}'=x_{+},\;x^{a'}=x^{a},
\label{e36} \\
\pi_{\pm}'&=&\pi_{\pm},\;\pi_{a}'=\pi_{a}. \nonumber
\end{eqnarray}

\noindent We change, in fact, only the coordinate  $x_{-}$, and therefore the
primes at the other variables are omitted below. In the new variables the
constraints surface is described by equations

\begin{equation}    \label{e37}
\Phi_{1}=\Phi^{G}=x_{-}'=0,\;\Phi_{2}=\Phi^{(1)}=\pi_{-}-\frac{\left[\pi_{a}
+eA_{a}(\zeta\tau)\right]^{2}+m^{2}}{\pi_{+}}=0.
\end{equation}

The Hamiltonian $H^{'(1)}$, arising as a result of the canonical transformation
(\ref{e36}), has the form

\[
H^{'(1)}=H^{(1)}+\frac{\partial W}{\partial\tau}=H^{(1)}+|\pi_{-}|,
\]

\noindent and on the constraint surface reduces to

\begin{equation} \label{e38}
H=|\pi_{-}|=\frac{\left[\pi_{a}+eA_{a}(\zeta\tau)\right]^{2}+m^{2}}{|\pi_{+}
|}.
\end{equation}

\noindent Because of one constraint depends on $\tau$, we have to solve all
the equations of quantization (\ref{e3}) - (\ref{e5}) or (\ref{e10}),
(\ref{e11}). The Hamiltonian (\ref{e38}) does not equal zero, so the
"Schr\"odinger" picture and Heisenberg picture differ one another. Let we
quantize in the "Schr\"odinger" picture. The nonzero Dirac brackets are the
same as in (\ref{e20}), (\ref{e21}). Thus, the time-equal commutation
relations for the operators  $(\hat{x}_{\pm}, \hat{x}^{a}, \hat{\pi}_{a})$
look like (\ref{e29}).
 On account of only the constraint $\Phi_{2}$  depends on time explicitly, and
the constraint $\Phi_{1}$  commutes with all independent variables and does not
only commute with  $\pi_{-}$,  no one independent operator depend on time,
according the eq. (\ref{e4}). The time evolution of the operator  $\hat{\pi}_
{-}$  is consequence of the equation of constraint   $\Phi_{2}=0$. The
Schr\"odinger equation has to be written in the form (\ref{e28}). Starting
that moment one can repeat all arguments of the previous (after eq. (\ref{e28})
 consideration, to demonstrate that, as a result of quantization, we get the
Klein-Gordon theory for a particle in a plane wave.

\section{Discussion}

First of all, we would like to underline the principle character of the
quantization rules (\ref{e2})-(\ref{e5}) in the case of time-dependent
constraints. Indeed, in that case the Schr\"odinger equation alone can not
determine the
time evolution of a quantum system. That evolution is determined both by the
equation (\ref{e4}) for the "Schr\"odinger" operators and by the Schr\"odinger
equation (\ref{e5}) itself. If the Hamiltonian $H$ is equal to zero on the
constraints surface, like, for instance, in the relativistic particle theory or
in Gravity, whole the time dependence is due to constraints, intrinsic ones and
gauge conditions. In this case the eq. (\ref{e4}) coincide with eq.(\ref{e10})
and fully determines the dynamics. This equation shows us what a freedom we
possess in forcing upon of the dynamics by means of time-dependent gauge
conditions.

As we said, in general case the dynamics is determined by both eq.(\ref{e4})
and the Schr\"odinger equation and is not unitary one. Nevertheless, there
exist particular cases (particular time-dependence of constraints) when whole
the dynamics,
or the dynamics in the physical sector only, is in fact unitary and can be
discribed in the frame of the Schr\"odinger equation only with some effective
Hamiltonian. The simplest one is the case when one can fulfil a canonical
transformation to new  variables in which the constraints do not already
depend on time. That means the time evolution due to the constraints
changing is a canonical transformation. As an example we can refer to the
problem of quantization of a relativistic particle in the constant magnetic
field in the time-dependent gauge of the form $x^{0}+\tau{\rm sign}\pi_{0} =0$,
where $\pi_{0}$ is the momentum conjugated to $x^{0}$ (see \cite{ab3}). The
example consdered in the present paper is more complicated because of the
above mentioned canonical transformation does not exist. But, as we have
demonstrated in item b) of Sect.3, there exists  such a canonical
transformation, that in new variables the dynamics in the physical sector
(here - the sector of independent, unconstrained variables) is unitary one
and can be only discribed by the Schr\"odinger equation with some effective
Hamiltonian. One can formulate conditions for constraints in general terms,
which correspond to the latter important case. So, let we have the theory with
the
equations of motion (\ref{e1}), and let $\eta'$ are some new  variables
connected with initial ones by means of a canonical transformation. We denote
the constraints in the new variables as $\Phi'(\eta',t)=0$ and the
corresponding Hamiltonian as $H'(\eta',t)$.The constraints, as any independent
set of second-class constraints \cite{ab2}, can always be solved explicitly
with respect to a part of the variables $\underline{\eta}'$,

\begin{equation} \label{e39}
\underline{\eta}'=f(\bar{\eta}',t), \; \eta'=(\underline{\eta},\bar{\eta}),
\end{equation}

\noindent so that $\underline{\eta}'$ and $\bar{\eta}'$ are sets of pairs of
canonically conjugated variables,
$\underline{\eta}'=(\underline{q}',\underline{p}'), \;
\bar{\eta}'=(\bar{q}',\bar{p}') $ .
The equations of motion (\ref{e1}) for the new variables can be
written  in terms of the equivalent set of constraints $\bar{\Phi}'$ and the
reduced Hamiltonian $\bar{H}'$ as follow \cite{ab2} :

\begin{eqnarray}
&&\dot{\bar{\eta}}'=\{\bar{\eta}',\bar{H}'+\epsilon\}_{D(\bar{\Phi}')}, \;
\bar{\Phi}'=\underline{\eta}'-f(\bar{\eta}',t)=0, \label{e40} \\
&& \bar{H}'=H'|_{\underline{\eta}'=f(\bar{\eta}',t)} \; . \nonumber
\end{eqnarray}

\noindent If for any $k$ and $m$ the equation holds,

\begin{equation} \label{e41}
\{\bar{\eta}'_{k},f^{l}\}\{\bar{\Phi}',\bar{\Phi}'\}^{-1}_{ll'}\{f^{l'},
\bar{\eta}'_{m}\}=0 \; ,
\end{equation}

\noindent then the Dirac bracket in (\ref{e40}) reduces to the Poisson
bracket and, because of $\{\bar{\eta}',\epsilon\}=0$ , the system (\ref{e40})
appears to be an ordinary system of Hamiltonian equations for the
independent variables $\bar{\eta}'$ ,

\[
\dot{\bar{\eta}}'=\{\bar{\eta}',\bar{H}'\} \; .
\]

\noindent In this case the quantization in the physical sector can be
done in the usual way by means of Poisson bracket, all the corresponding
operators being time-independent in the Schr\"odinger picture. The dynamics
is unitary one and is controled by the Schr\"odinger equation with the
Hamiltonian $\bar{H}'$. That can also be seen  from the eq. (\ref{e4}),
which reduces in that particular case to $\dot{\hat{\eta}}'=0$. One can point
out the possible structure of the functions $f(\bar{\eta}',t)$ from eq.
(\ref{e39}), which obey   the condition (\ref{e41}). Namely, if for each pair
of conjugated variables $\underline{\eta}'^{l}=(\underline{q}'^{n},\underline
{p}'_{n}), \; l=(\zeta,n), \;\zeta=1,2$, at least one of the function
$f^{1n},f^{2n}$ in (\ref{e39}) be equal to zero (so called constraints of
special form \cite{ab2}), then the condition (\ref{e41}) holds. In the example
 considered in the Sect.3, the constrains (\ref{e37}) have just the above
mentioned special form.

Finally, in the connection with the concrete problem of the quantization of
the relativistic particle in the presence of various backgrounds, one ougtht
to say the follow. If one uses the method of quantization in which all
first-class constraints are considered in the sense of restrictions on the
state
vectors, and the commutation relations are  built by means of the Dirac
bracket taken with respect to second-class constraints, without the adding of
any gauge conditions on classical level, then the Klein-Gordon equation or
Dirac equation formally appear immediately, after the transition from
classical variables to the corresponding operators, because of some of
first-class consraints have just the form of these equations. Unfortunately,
in
this approach the problem of the Hilbert space definition stays unsolved in
general case. In the frame of canonical operator quantization, which is, in
fact, quantization in the physical sector, and which we have applied here and
in
\cite{ab3}, the problem looks technically more complicated, but all the
steps are consistent and grounded. Nevertheless, on such a way each concrete
background needs to be analysed especially. As to electromagnetic backgrounds,
they can be divided in   three principal classes, within each of them
the quantization problem looks similar to our mind. The simplest, in that
sense,
 is the class of constant electromagnetic fields, which, at the same time,
does not violate the vacuum stability (does not create pairs from vacuum)
from the point of view of the QED with an external field \cite{ab7}. We have
demonstrated the quantization in this case on the example of a relativistic
particle in the constant magnetic field \cite{ab3}. In all such fields there
exists a discrete integral of motion, which gives the classification in
particles and antiparticles. Besides, there exists such a chronological gauge
in which all the dynamics is unitary one. To the second class belong
electromagnetic fields, which can depend on time, but do not violate the vacuum
 stability. The typical configuration here is a plane wave field, the
quantization in this background was considered in the present paper. In all
fields, belonging to that class, as before, one can find a discrete integral
of motion (in our case $\zeta$),
 which gives the classification in particles and antiparticles. As to
dynamics, one can choose such a chronological gauge in which the dynamics in
physical sector stays unitary. To the third class belong external
electromagnetic fields, which do violate the vacuum stability. The typical
representative here is an electric field. The difficulties which appear in
this case are connected with the lack of the above mentioned integral of
motion, which gives the classification in particles and antiparticles. The
same situation  exists in an external gravitational field.
We believe the
solution of the problem in the case of an electromagnetic field violating the
vacuum stability can give the key to the  canonical quantization of the
relativistic particle
in gravitational background.


\begin{thebibliography} {9}
\bibitem{ab1}P.A.M.Dirac, {\em Lectures on Quantum Mechanics} (Befer Graduate
School of Science, Yeshiva University, New York 1964)
\bibitem{ab2}D.M.Gitman, I.V.Tyutin, {\em Quantization of Fields with
Constraints} (Springer-Verlag, 1990)
\bibitem{ab3}D.M.Gitman, I.V.Tyutin, Class. Quantum Grav. {\bf 7} (1990) 2131
\bibitem{ab4}V.G.Bagrov, D.M.Gitman, {\em Exact Solutions of the Relativistic
Wave Equations} (Kluwer, 1990)
\bibitem{ab5}R.P.Feynman, Phys.Rev. {\bf 76} (1949) 749
\bibitem{ab6}S.P.Gavrilov, D.M.Gitman, Sh.M.Shvartsman, Yadern. Fiz. {\bf 29}
 (1979) 1392
\bibitem{ab7}E.S.Fradkin, D.M.Gitman, Sh.M.Shvartsman, {\em Quantum
Electrodynamics with Unstable Vacuum} (Springer-Verlag, 1991)
\end{thebibliography}
\end{document}